\documentclass[8pt,aps]{revtex4}
\usepackage{amssymb}
\usepackage{latexsym}

\usepackage{epsfig}

\usepackage{dcolumn}
\usepackage{amsmath}
\usepackage{amsfonts}
\usepackage{bm}

\newcommand{\be}{\begin{equation}}
\newcommand{\ee}{\end{equation}}
\newcommand{\bea}{\begin{eqnarray}}
\newcommand{\eea}{\end{eqnarray}}

\newcommand{\p}{\partial}
\newcommand{\s}{\sigma}

\newcommand{\la}{\langle}
\newcommand{\ra}{\rangle}
\newcommand{\rd}{\mbox{d}}
\newcommand{\ri}{\mbox{i}}
\newcommand{\re}{\mbox{e}}

\renewcommand{\vec}[1]{{\bm #1}}

\begin{document}
\title{An analytically tractable  model of bad metals}
\author{S. Akhanjee and A.M. Tsvelik}
\affiliation{ Department of Condensed Matter Physics and Materials Science, Brookhaven National Laboratory,
  Upton, NY 11973-5000, USA} \date{\today } 
\begin{abstract} 
We discuss a model Kondo-type Hamiltonian representing an analytically tractable version of the model used by Yin {\it et.al.}, Phys. Rev. B{\bf 86}, 2399 (2012) to explain the non-Fermi liquid behavior of iron chalcogenides and ruthenates in an intermediate energy range.  We consider a regime where a complete screening of the local degrees of freedom proceeds in two stages described by two characteristic energy scales $T_K^{orb} >> E_0$. The first scale marks a screening of the orbital degrees of freedom and the second one marks a crossover to the regime  with coherent propagation of quasiparticles. We present analytical results for the specific heat and magnetic susceptibility at $T << T_K^{orb}$. 

 \end{abstract}


\maketitle
\section{Introduction}

There is a significant number of  metallic systems dubbed "bad metals'' where the energies at which quasiparticles emerge as coherent objects  are much lower than the characteristic scale of the interactions. In \cite{millis},\cite{gabi} it was suggested  that in the compounds based on $d$ and $f$-elements,  the Hund's interaction plays major role in delaying the onset of coherence. Those authors considered models where electrons carry both orbital and spin indices using the  Dynamical Mean Field Theory (DMFT) and  found that the coherence scale $E_0$ was indeed low in comparison with the bandwidth $W$ or the Hund's rule coupling $J_H$. In the intermediate range $E_0 < T,\omega < $min$(W, J_H)$ the electronic self energy was found to have a non-Fermi liquid form $\Sigma \sim \omega^b$. In \cite{millis} the  exponent was found to be  universal $b =1/2$, in \cite{gabi} which used a more realistic model $b$ was non-universal. The latter result was found to fit the observed behavior of the mid-infrared optical conductivity in the iron chalcogenides and ruthenates. 

 The starting point for \cite{gabi} is the Kondo lattice model where the localized $d$-electrons (6 electrons per site) give rise to spin and orbital moments: 
\bea
&& H_K =\sum_k \epsilon(k) \psi_{j\s}^+(k)\psi_{j\s}(k) +\label{model1}\\ 
&& \frac{1}{N}\sum_{{\bf r}}\re^{\ri {\bf qr}}\Big[ J_1\psi^+_{j\s}(q+k)X_{lj}({\bf r})\psi_{l\s}(k)  + J_2\psi^+_{j\s}(q+k)[X_{lj}\vec S]({\bf r}){\vec\s}_{\s\s'}\psi_{l\s'}(k) - J_3\psi^+_{j\s}(k+q)\Big({\vec \s}_{\s\s'}{\vec S}({\bf r})\Big)\psi_{j\s'}(k)\Big],  \nonumber
\eea
where $N$ is the number of sites. The  spin degrees of freedom are described by spin $S=2$ operators acting on the spin indices and the orbital sector is described by the Hubbard operators $X_{jl}$ (j, l =1,...M). The symmetry of Hamiltonian (\ref{model1}) is SU(M)$\times$SU(2)$\times$U(1).  In iron pnictides and chalcogenides, a Fe-ion is surrounded by a tetrahedron of pnictogen or chalcogen, and the resulting crystal field is weak in comparison to the Fe-pnictogen hybridization. As a result all $d$ orbitals  have degeneracy, $M=5$. In the ruthenates where the coordination of Ru-ion is octohedral the crystal field is strong, yielding $M=3$ because only the $t_{2g}$ orbital actively participates in the interaction.  A detailed description of the model can be found in the Supplementary material to \cite{gabi}. Furthermore, a similar model with different parameters was considered in \cite{piers}. 

 As we have mentioned above, model (\ref{model1}) was treated by DMFT \cite{gabi}. In  that approach  local moments from different sites have been considered as uncorrelated. As a result the Kondo lattice was treated as a  single impurity problem with a self-consistently renormalized  density of states (DOS) of the band electrons.  It turned out, however, that in the given case the DOS at the chemical potential remains non-singular and hence the self-consistency had no qualitative effect. Therefore  the range of energies where the magnetic interactions between sites are still small can be treated as a single impurity problem.

Additionally, the DMFT treatment established the  existence of the intermediate regime marked by nontrivial power laws in the electron self energy. In all likelyhood  this regime emerges as a crossover between the quantum critical point (QCP) of the purely orbital Kondo model ($J_{2} =0$) and the Fermi liquid strong coupling regime of the full model ($J_2 \neq 0$). Indeed, at $J_2=0$ the single impurity Kondo model decouples into two independent Kondo models describing scattering in the orbital and the spin  sectors. This decoupling occurs due to the fact that  electronic densities ("currents'') coupled to the  orbital $X_{jl}$ and spin operators $S^a$ commute with each other. Both spin and orbital Kondo models are overscreened and at strong Hund's coupling, the sign of the exchange interaction in the spin channel is ferromagnetic \cite{gabi}. Therefore, at $J_2=0$ the orbital Kondo model scales to the QCP  characterized by nontrivial exponents and the spin one scales to weak coupling. However, as soon as the coupling  $J_2$ is switched on, the renormalization group trajectories start to deviate from the critical point. Eventually the system becomes a Fermi liquid. The corresponding energy scale  is, of course, determined by $J_2$.

 Although the meaning of  the numerical results by \cite{gabi} is sufficiently transparent, it would be advantageous to have a simple model of bad metals where a qualitatively similar picture can be obtained analytically.  Therefore,  this is the primary focus of the present paper, which is organized as follows: The model is described in Section II; in the same section we present its solution and describe the low temperature thermodynamics. In Section III we briefly discuss the general case (that is the single impurity version of model (\ref{model1})). Lastly, we provide an overview of our results in the conclusion.  

\section{A solvable single impurity model} 

Below  we consider a single impurity version of model (\ref{model1}) where an analytical treatment is possible. This version has four species of fermions with symmetry SU(2)$\times$SU(2)$\times$U(1) interacting with pseudospin ${\bf T}$ and spin ${\bf S}$, both of magnitude 1/2. The interaction has a form:
\bea
V =  g_1\psi^+(I\otimes\tau^a)\psi \hat T^a +   2g_2\psi^+(\s^a\otimes\tau^b)\psi\hat S^a\hat T^b + g_3\psi^+(\s^a\otimes I)\psi \hat S^a \label{V}
\eea
The Pauli matrices $\s^a$ and $\tau^a$ act in the spin and the orbital sectors respectively,  and obey the identites,
\bea
\s^a\hat S^a -1/2 = \hat P_{spin}, ~~ \tau^a \hat T^a +1/2 =\hat P_{orb}, 
\eea
where $P_{spin}, P_{orb}$ are  permutation operators acting in the corresponding sectors. Therefore, when $g_1 = g_2 = g_3$ the interactions   become $\hat P_{spin}\hat P_{orb} = \hat P_{SU(4)}$,  which yields the integrable SU(4) Coqblin-Schrieffer model\cite{wiegmann}. At low temperatures the latter model displays  Fermi liquid behavior characterized by the complete screening of the impurity. When the couplings are not equal to each other the model is not integrable (except for the case $g_2 =0$), but in all likelyhood eventually reaches the regime of full screening.  

 We will be interested in the case  $g_1 >> g_{2,3}$ when the orbital sector reaches the critical point first. As is well known, for the single impurity problem, the dimensionality of the bulk is irrelevant as  long as the density of states (DOS)  at the chemical potential is constant. It allows one to treat the bulk in the single impurity Kondo problem as a one-dimensional theory of chiral fermions. Such  a replacement carries many advantages enabling one to apply various non-perturbative techniques available for one-dimensional theories, such as Bethe ansatz and non-Abelian bosonization described in \cite{wznw},\cite{wznw1}. In  the latter procedure one replaces  the bulk Lagrangian of 1D fermions with U(1)$\times$SU(2)$\times$SU(2) symmetry  by an equivalent representation consisting of a U(1) bosonic theory describing charge fluctuations and two SU$_2$(2) Wess-Zumino-Novikov-Witten (WZNW) Lagrangians describing the spin and the orbital sectors. The peculiarity of the $N=M=2$ case is that the WZNW models  are equivalent to models of Majorana fermions. The net result is:
\bea
&&\sum_k  \psi^+_{j\s}(k)(\p_{\tau} - \epsilon_k)\psi_{j\s}(k) = \label{decomp}\\
&& \int_{-\infty}^{\infty} \rd x \Big[\p_x\phi(\ri\p_{\tau} +\p_x)\phi + \frac{1}{2}\chi_a(\p_{\tau}- \ri\p_x)\chi_a + \frac{1}{2}\xi_a(\p_{\tau} - \ri\p_x)\xi_a\Big],\nonumber
\eea
where $\xi_a,\chi_a$ ($a=1,2,3$) are Majorana fermions transforming according to the adjoint representation of the SU(2) group. The scalar bosonic field $\phi$ describes the charge sector which is decoupled from the impurity. Henceforth, we set the Fermi velocity $v_F =1$ and thus $\rho(\epsilon_F) = 1/2\pi$. The total central charges of the left- and right-hand side of (\ref{decomp}) are, naturally, equal: 4= 1 + 3/2 + 3/2. 
 It follows that the electron current operators can be expressed as 
\bea
&& \psi^+{\tau^a}\psi \rightarrow \frac{\ri}{2}\epsilon^{abc}\chi_b\chi_c, ~~ \psi^+\s^a \psi \rightarrow \frac{\ri}{2}\epsilon^{abc}\xi_b\xi_c\label{curr1}\\
&& \psi^+{\tau^a\s^b}\psi \rightarrow \ri \chi^a\xi^b \label{curr2}
\eea
As was shown in \cite{ioffe}, at the 2-channel Kondo model QCP the local pseudospin $T^a$ renormalizes into \cite{ioffe}
\bea
\hat T^a \rightarrow {T_K^{orb}}^{-1/2}\epsilon \chi_a(0) \label{spin}
\eea
where $\epsilon$ is a local zero energy Majorana mode residing on the impurity site $x=0$ and $T_K^{orb} \sim Wg_1\exp(- \pi/g_1)$ is the orbital Kondo temperature. 

As a result we obtain the following  effective Lagrangian describing the behavior at energies below $T_K^{orb}$ (the U(1) part of the bulk Lagrangian is omitted):  
\bea
&& L = g_2^*(T_K^{orb})^{1/2}[\xi^b(0) \hat S^b]\epsilon  + \ri \frac{g_3^*}{2}\epsilon_{abc}\hat S^a\xi^b(0)\xi^c(0) + \frac{1}{2}\int \rd x \xi_a(\p_{\tau}- \ri\p_x)\xi_a \nonumber\\
&& + \Big[\frac{1}{2}\int \rd x \chi_a(\p_{\tau} - \ri\p_x)\chi_a + \frac{1}{2}\epsilon\p_{\tau}\epsilon + {T_K^{orb}}^{-1/2}\epsilon \chi_1(0)\chi_2(0)\chi_3(0)\Big], \label{last}
\eea
where $g_2^*,g_3^*$ are renormalized values of the corresponding coupling constants. The term in the square brackets is the critical point Lagrangian of the orbital 2-channel Kondo model. The last operator in the square brackets is irrelevant, but it is included since it determines the impurity  contribution to the specific heat \cite{tsvelik}: 
\be
C_V \sim (T/T_K^{orb})\ln(T_K^{orb}/T). 
\ee
The most important part of (\ref{last}) is the first term,  which follows from the $g_2$ term in Eq.(\ref{V}) where we replaced the fermionic  bilinear  by (\ref{curr2}), the orbital operator $T^a$ by (\ref{spin}) and  applied the fusion rule:
\bea
2g_2\psi^+(\s^a\otimes\tau^b)\psi\hat S^a\hat T^b \rightarrow g_2^* (T_K^{orb})^{-1/2}[\chi^b\xi^a](t+\eta)[\hat S^a\chi^b](t)\epsilon(t) \approx  \frac{g_2^*}{\sqrt{T_K^{orb}} 2\pi \eta}[\xi^a\hat S^a]\epsilon\label{scr}
\eea
Since the interaction becomes retarded in the process of renormalization,  we have assumed that the operators should be time split by the amount $\eta \sim 1/T_K^{orb}$. As a result, at the QCP the operator in question  becomes a relevant one \cite{differ}. 
 
To solve model (\ref{last}) we  introduce  a transformation 
\be
 f^a = 2\ri\epsilon \hat S^a,
\ee
where $f^a$ are Majorana fermions, satisfying the Clifford algebra
\be
[f^a,f^b]_+ = 2\delta_{ab}.
\ee
This transformation respects the Casimir operator ${\bf S}^2 = 3/4$. 
Since $\epsilon^2 = 1$, the inverse transformation is 
\bea
\hat S^a = \frac{\ri}{4}\epsilon_{abc}f^bf^c, ~~ \epsilon = f_1f_2f_3.\label{spin2}
\eea
The latter expression  reproduces the commutation relations of spin operators. It is essential that $f^a$'s replace completely $\epsilon, S^a$ and therefore field $\epsilon$ is no longer in use. 
Then  the effective Lagrangian (\ref{last}) becomes 
\bea
&& L = \ri g_2^*\sqrt{T_K^{orb}}[\xi^b(0) f^b]   + \frac{1}{2}\int \rd x \xi_a(\p_{\tau}- \ri\p_x)\xi_a + \frac{1}{2}f^a\p_{\tau}f^a \label{L1}\\
&& + \Big[\frac{1}{2}\int \rd x \chi_a(\p_{\tau} - \ri\p_x)\chi_a +\frac{g_3}{2}[f^b\xi^b(0)][f^c\xi^c(0)] + {T_K^{orb}}^{-1/2}f_1f_2f_3 \chi_1(0)\chi_2(0)\chi_3(0)\Big] \label{last1}.
\eea
The impurity thermodynamics is determined by the Green's functions of the $f$-operators. In the zeroeth order in $g_3$ and ${T_K^{orb}}^{-1/2}$,  one obtains these functions by diagonalizing the quadratic part  of the Lagrangian given by (\ref{L1}). The most convenient way to proceed is to write down the Lagrangian for the Fourier components of $f$ and $\xi(0)$:
\bea
L_0 = \sum_{\omega}\Big[ \frac{1}{2}\xi^a(-\omega,0)G_0^{-1}(\omega,x=0)\xi^a(\omega,0) + \ri \sqrt{E_0/2}\xi^a(-\omega)f^a(\omega) + \frac{\ri\omega}{2}f^a(-\omega)f^a(\omega)\Big]
\eea
where 
\[
G_0(\omega,x=0) = \int \frac{\rd k}{2\pi} \frac{1}{\ri\omega - k} = -\frac{\ri}{2}\mbox{sign}{\bf (\omega)}
\]
 is the Green's function of the bulk fermion at $x=0$ and $E_0 = \frac{1}{2}[g_2^*]^2T_K^{orb}$.  The net result is  
\bea
\left(
\begin{array}{cc}
\la\la f(\omega)f(-\omega)\ra\ra & \la\la f(\omega)\xi(-\omega,0)\ra\ra\\
\la\la\xi(\omega,0) f(-\omega)\ra\ra & \la\la\xi(\omega,0) \xi(-\omega,0)\ra\ra
\end{array}
\right) = -\frac{1}{|\omega_n| + E_0}\left(
\begin{array}{cc}
\ri\mbox{sign}\omega_n & -\ri\sqrt {E_0/2}\\
\ri\sqrt{E_0/2} & \ri\omega_n/2
\end{array}
\right), \label{spin3}
\eea
 From the form of (\ref{spin3}) is clear that $E_0$  is the crossover energy scale to the Fermi liquid regime. 

The nonquadratic terms in (\ref{last1}) provide subleading corrections to the scaling and hence are irrelevant in the infrared. The corresponding corrections  can be calculated by the perturbation theory using Green's functions (\ref{spin3}). For instance,  the  ferromagnetic exchange interaction $g_3$  reduces the value of $E_0$ so that the new scale $E_{eff}$ is given by:
\bea
&& E_{eff}^{1/2} = E_0^{1/2} - 2g_3\sqrt 2\ri T\sum_{\omega}\la f(\omega)\xi(-\omega,0)\ra, \nonumber\\
&& E_{eff}^{1/2} \approx \frac{E_0^{1/2}}{1 -  g_3/\pi \ln[T_K^{orb}/\mbox{max}(T,E_{eff})]}. \label{new}
\eea
 The onset of Fermi liquid is delayed till the new scale $E^*$ determined by $T=0$ limit of equation (\ref{new}):
\be
E^*[1- \frac{g_3}{\pi}\ln(T_K^{orb}/E^*)]^2 = E_0
\ee
which can be significantly smaller than $E_0$ if the ferromagnetic exchange is strong. Otherwise the weak temperature dependence of $E_{eff}$ will give logarithmic corrections to the scaling considered below. In the first approximation the thermodynamics is still determined by (\ref{magn},\ref{C}) with $E_0$ replaced by $E_{eff}(T)$ determined by (\ref{new}).  
The last term in (\ref{last1}) containing six fermionic operators is even less singular at $T << T_K^{orb}$.  

 In the first approximation one can neglect the temperature dependence of $E_{eff}$ given by (\ref{new}) and replace it by $E_0$. The temperature dependence of the magnetic susceptibility (displayed on Fig. 1) is given by the formula \cite{GR}
\bea
&& <<\hat S^z \hat S^z>> = \chi(T,\omega =0) = \frac{2}{\pi}\int_0^{\infty}\frac{\omega E_0}{(\omega^2 +E_0^2)^2}\tanh(\omega/2T)\rd\omega \nonumber\\
&& = \frac{1}{\pi  E_0}\Big[1 - (2\pi T/E_0)^2 + \frac{2\pi T}{E_0}\psi'\Big(E_0/2\pi T + 1/2\Big)\Big]\label{magn},
\eea
which at $T > 0.3 E_0$  can be approximated with high accuracy  as
\bea
\chi(T) \approx \frac{1}{4(T + 0.59 E_0)}
\eea
\begin{figure}
\centerline{\includegraphics[angle = 0,
width=0.5\columnwidth]{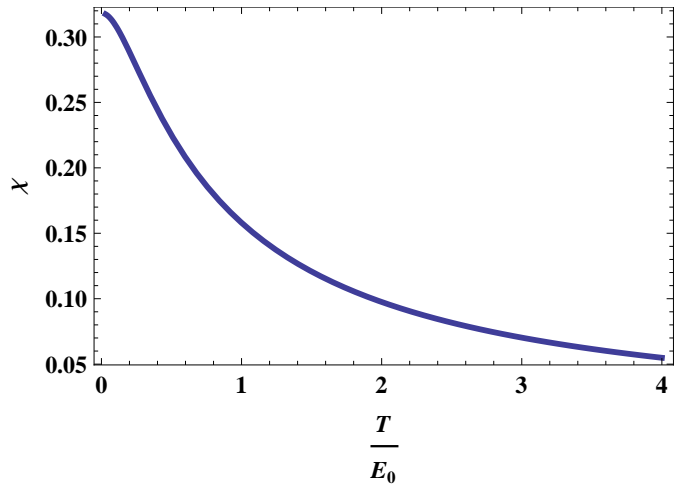}}
\caption{Temperature dependence of the magnetic susceptibility (\ref{magn}).}
 \label{pict1}
\end{figure}
The spin sector contribution to the specific heat is 
\bea
C = \frac{3T}{4\pi E_0}\int_0^{\infty}\frac{\rd x x^2}{[x^2(T/E_0)^2 +1]\cosh^2(x/2)} = \frac{3T}{4\pi E_0}\Big[1- \frac{E_0}{2\pi T}\psi'\Big(E_0/2\pi T + 1/2\Big)\Big]. \label{C}
\eea

\begin{figure}
\centerline{\includegraphics[angle = 0,
width=0.5\columnwidth]{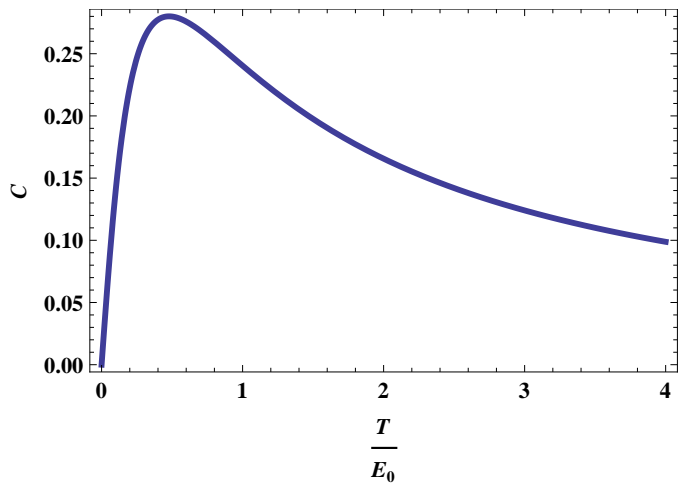}}
\caption{Temperature dependence of the spin sector contribution to the specific heat (\ref{C}).}
 \label{pict2}
\end{figure}
We see that at $T << E_0$ the susceptibility is constant and the specific heat is linear in T. Hence at energies smaller  than $E_0$ the impurity spin is fully screened.

\subsection{Spin-orbit coupling}

 Our model (\ref{V}) can be easily augmented by inclusion of the spin-orbit coupling:
\be
H_{SO} = \Lambda {\bf T}{\bf S}.
\ee
At the 2-channel QCP we replace ${\bf T}$ as in (\ref{spin}) to get
\bea
H_{SO} \rightarrow \ri\frac{\Lambda}{\sqrt {T_K^{orb}}}\epsilon\chi^aS^a
\eea
This term combines with (\ref{scr}) such that now the impurity spin interacts with a linear combination of the orbital and spin Majoranas:
\bea
\cos\alpha \chi^a + \sin\alpha \xi^a, ~~ \tan\alpha = g_2^*T_K^{orb}/\Lambda,
\eea
with the coupling constant 
\be
\tilde V = (\Lambda^2/T_K^{orb} + (g_2^*)^2T_K^{orb})^{1/2}.
\ee
Thus, we see that the spin-orbit coupling does not bring any qualitative changes.

\section{A model with generic symmetry}

As  mentioned earlier, a single impurity version of model (\ref{model1}) is not solvable. However,  it is still possible and  worthwhile to make  some qualitative statements about it. Consider  model (\ref{model1}) with U(1)$\times$SU(N)$\times$SU(M) symmetry. An analog of decomposition (\ref{decomp}) in this case is (see, for example, \cite{wznw},\cite{wznw1},\cite{ludwig}):
\bea
&&\sum_{j=1}^M\sum_{\s=1}^N\sum_k  \psi^+_{j\s}(k)(\p_{\tau} - \epsilon_k)\psi_{j\s}(k) = \label{decomp2}\\
&& \int_{-\infty}^{\infty} \rd x \p_x\phi(\ri\p_{\tau} +\p_x)\phi + \frac{2\pi}{N+M}\int\rd x\Big[:J^aJ^a: + :F^aF^a:\Big],
\nonumber
\eea
where $J^a$ ($a= 1,... N^2-1)$ and $F^a$ ($a= 1,...M^2-1$) are  currents from SU(N)$_M$ and SU(M)$_N$ Kac-Moody algebras defined as 
\bea
J^a = \sum_{j=1}^M\psi^+_{j\alpha}\s^a_{\alpha\beta}\psi_{j\beta}, ~~ F^a = \sum_{\s=1}^N\psi^+_{j\alpha}\tau^a_{jk}\psi_{k\alpha}
\eea
with $\s^a,\tau^a$ being generators of the corresponding groups. For general $N \neq M$ the quadratic forms of currents cannot be expressed as models of free fermions as it happens for $N= M$, where they can be written in terms of Majorana fermions transforming in the adjoint representation of the SU(N) group. 
 Neither does it work for the cross term in (\ref{V}). Instead we have 
\bea
\psi^+{\tau^a\s^b}\psi \rightarrow \Phi^a\Lambda^b,
\eea
where $\Phi,\Lambda$ are Wess-Zumino primary fields from the adjoint representations of the SU(M) and SU(N) groups respectively. Their conformal dimensions are 
\bea
h_{\Phi} = \frac{M}{N+M}, ~~ h_{\Lambda}= \frac{N}{N+M}.
\eea
The QCP corresponds to the overscreening of the SU(N) sector. The corresponding Kondo scale is $T_K^{orb} = W g_1^{M/N}\exp(- 2\pi/Ng_1)$. At the QCP one operator $\Phi$ creates an average with the impurity spin leaving behind a zero mode,  fashioned along the lines of (\ref{spin},\ref{scr}).  As a result, the operator perturbing the critical SU$_N$(M) WZNW model is $\Lambda^b$ coupled to the SU(M) impurity spin. Its scaling dimension is $N/(N+M) < 1$ and hence it is relevant. Thus, we can be confident that the model scales to strong coupling, which is presumably a Fermi liquid, but  the details of this are not known. Consequently,  the corresponding energy scale marking the crossover to Fermi liquid is 
\be
E_0 \sim g_2^{1/(1-h_{\Lambda})}T_K^{orb}
\ee
\section{Conclusion and Acknowledgements}

 We have presented an analytically tractable  single impurity Kondo model where the impurity carries both spin and pseudospin $S=T$ =1/2 and the bulk electrons carry both spin and orbital indices. Our original motivation for considering this model was its relevance to the problem of "bad" metals, as formulated in \cite{millis},\cite{gabi},\cite{piers}. As we have already mentioned, according to DMFT, the single impurity problem describes both thermodynamics and the  electron self energy within a range where the intersite correlations are relatively weak.  Although the experimentally relevant models have different  values of $S$, we have chosen $S=1/2$ to obtain analytic results. 

  Furthermore, we have considered the most interesting regime when the orbital moment is screened first and the coupling between spin and orbital channels is weak. Therefore, the screening of the orbital moment leads to  a non-Fermi-liquid quantum critical point (QCP) which is destabilized by the interaction term mixing spin and orbital channels. If the corresponding coupling constant $g_2$ is small, there is a wide crossover regime between  the unstable two-channel Kondo model QCP  at high energies and coherent Fermi liquid at low energies.  The ratio of these two energy scales is $\sim g_2^2$. 

 It is quite likely that the nonuniversal exponents  in the self energy \cite{gabi} discussed in the Introduction are crossover effects. To illustrate this point one needs to calculate the self energy; we plan to do it in  a subsequent paper.

 AMT is grateful to G. Kotliar, S. Lukyanov, R. M. Konik and P. Coleman for inspirational discussions and to A. James for help with the numerical calculations. The work was supported by the US DOE under contract number DE-AC02-98 CH 10886.


\begin{thebibliography}{99}
\bibitem{millis} P. Werner, E. Gull, M. Troyer, and A. J. Millis, Phys. Rev. Lett. {\bf 101}, 166405 (2008). 
\bibitem{gabi} Z. P. Yin, K. Haule, G. Kotliar, Phys. Rev. B{\bf 86}, 239904(E) (2012).
\bibitem{piers} T. T. Ong, P. Coleman, Phys. Rev. Lett. {\bf 108}, 107201 (2012). 
\bibitem{wiegmann}  A. M. Tsvelick and P. B. Wiegmann, J. Phys. C{\bf 15}, 1707 (1982).
\bibitem{wznw} A. M. Tsvelik, "Quantum Field Theory in Condensed Matter Physics", Ch. 31,32; Cambridge University Press, 2003. 
\bibitem{wznw1} P. Di Francesco, P. Mathieu, D. Senechal, "Conformal Field Theory", Springer 1996. 
\bibitem{maldacena} J. M. Maldacena and A. W. W. Ludwig, Nucl. Phys. B{\bf 506}, 565 (1997).
\bibitem{ioffe} P. Coleman, L. B. Ioffe and A. M. Tsvelik, Phys. Rev. B{\bf 52}, 6611 (1995).
\bibitem{tsvelik} A. M. Tsvelick, J. Phys. C{\bf 18}, 159 (1985). 
\bibitem{differ} From this point on our results differ from the ones of \cite{piers}. 
\bibitem{GR} I. S. Gradshtein and I. M. Ryzhik, "Table of Integrals, Series and Products", sixth edition, 3.415 (4).  
\bibitem{ludwig} I. Affleck, A. W. W. Ludwig, Phys. Rev. B {\bf 48}, 7297 (1993). 
\end{thebibliography}
\end{document}